# Freeform Spectrally Stable Topological Photonic Vortex Resonators

Yuma Kawaguchi[1]*, Daria Smirnova[2]*, Filipp Komissarenko[1,3], Daria Kafeeva[1], Svetlana Kiriushechkina[1], Jeffery Allen[4], Monica Allen[4], Andrea Alù[1,5,6], and Alexander Khanikaev[3]

[1]Department of Electrical Engineering, The City College of New York, New York, NY 10031, USA
[2]Research School of Physics, The Australian National University, Canberra, ACT 2601, Australia
[3]College of Optics and Photonics – CREOL, University of Central Florida, Orlando, FL 32816, USA
[4]Air Force Research Laboratory, Munitions Directorate, Eglin AFB, Eglin, FL 32542, USA
[5]Photonics Initiative, Advanced Science Research Center, City University of New York, New York, NY 10031, USA
[6]Physics Program, Graduate Center, City University of New York, New York, NY 10016, USA
*These authors contributed equally to the present manuscript

**Abstract.** Topological concepts have been at the forefront of materials research in recent years, driving a revolution in our understanding of the response of quantum materials and enabling new ways to manipulate light and sound in topological metamaterials. Topological defects and topological boundaries of different dimensions have driven a paradigm shift in photonics, where topological photonic crystals and metamaterials can be engineered to create one-way flow of energy robust to defects or to control such flows with synthetic degrees of freedom along topological domain walls. More recently, topological point singularities encoded into photonic structures have been shown to enable confinement of optical modes with the topologically nontrivial nature of the cavity imprinted into the vorticity of optical far fields. Here we demonstrate that the two latter concepts - domain wall and point singularities - can be unified into an even more powerful tool to enable arbitrarily shaped resonant cavities of any dimension supporting spectrally stable "zero-energy" modes. We experimentally confirm that such modes, whose existence is guaranteed by topological principles, allow an unprecedented degree of control over the optical field, which appears to have no phase modulation across space, can have any desirable radiation pattern, and enables spectral stability regardless of shape or length.

## Introduction

Designer materials with highly tailorable characteristics across physical domains are driving a new technological revolution in front of our eyes. Designer photonic materials, such as photonic crystals and metamaterials, have been supporting an unprecedented variety of applications across the electromagnetic spectrum[1,2]. Topological materials represent a recent family of designer materials with unique properties, stemming from deliberately engineered topologically nontrivial structure of their modes[3-6]. Among the most fascinating properties enabled by the topological nature of these materials are the inherent robustness of the modes to defects and disorder[7], their unidirectional propagation[8,9], and the opportunity to control modes via synthetic degrees of freedom[10].

Topological photonics has become one of the most fruitful playgrounds for exploiting these properties, from robust waveguides and resonators, to nonlinear effects, strong light-matter interaction, and lasers, all leveraging topological modes of different nature and dimensionality[11-15]. As of today, topological guided states of valley and spin-Hall type are arguably the most used due to the well-established design approach based on "photonic graphene" in photonic crystals and its various perturbations, which allow attaining valley and pseudo-spin polarized transports across frequencies[16-23]. Another, more recent but quite powerful approach is based on topological vortex resonators, which offer modes trapped to a point defect – a vortex – with a controllable degree of localization around it[24-31]. Notably, such resonators can be implemented in the same kind of systems as guided boundary modes.

Here we demonstrate that these two seemingly distinct topological platforms can in fact be combined to enable a new, more versatile tool for generating topological resonances. Unlike other resonators based solely on domain walls[19,32-34], these topological resonators always exhibit at least one mode pinned to the "zero energy" and thus invariant with respect to variations in length or shape of the cavity. We experimentally confirm spectral stability for both 1D and 2D resonators of different length and shape, also demonstrating the possibility to control the far-field radiation patterns from such resonators.

In the context of our findings, relevant recent studies have shown that wave propagation can be made quasi-static and with a uniform phase over long distances in epsilon-near-zero (ENZ) media[35] or operating at the $\Gamma$ point of a photonic crystal[36]. By building zero-th order resonators in such materials, resonances largely independent of geometry can emerge[37,38]. Consistent with this interest in flexible, free-form resonant systems, our approach introduces a novel route to free-form resonances that combines design freedom with inherent topological robustness.

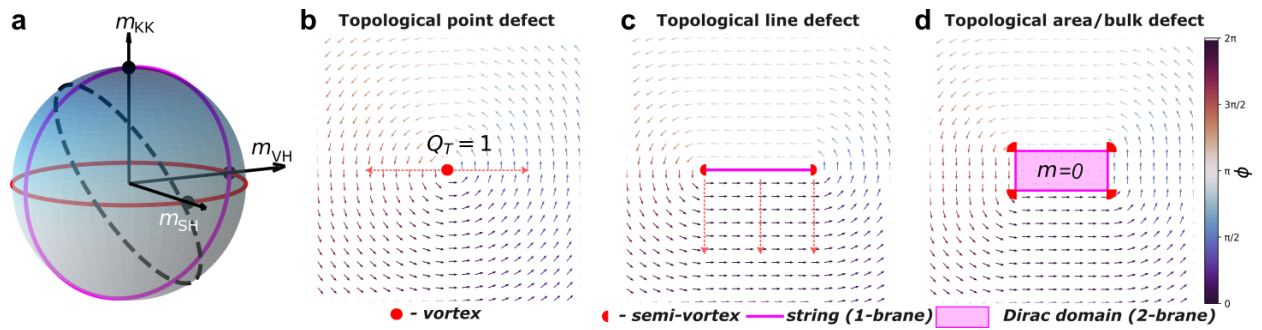

**Fig. 1. A cross-dimensional family of topological vortex defects. a**, A sphere of masses due to three different types of perturbations, spin-Hall, valley-Hall, and Kekulé, acting independently in synthetic Hilbert space of the lattice. **b**, Topological vortex defect with the topological charge (winding) $Q_T = 1$, and its transformation into a line defect formed by a domain wall terminated by semi-vortexes in **c** and into a bulk $m = 0$ Dirac (gapless) cavity with winding in the outer domain in **d**. In all three cases the winding guarantees the presence of "zero-energy" mode whose frequency is independent of the size of the cavity.

**Theoretical background**

The conceptual idea of freeform topological vortex resonators is summarized in Fig. 1. The core photonic system used here represents a variation of 2D topological photonic crystal with the Dirac spectrum (with our final design based on Hafezi's[17] proposal – a graphene lattice of triangles). Perturbation in a photonic Dirac system allows three distinct gapped phases (i) valley-Hall type by dimerization of graphene lattice atoms/sites, (ii) Wu and Hu (pseudo-)spin-Hall type by shrinkage/expansion of nearby hexamers of atoms, and (iii) what we refer to as Kekulé type attained by a trimerization of a hexamer by bringing pairs of atoms closer/farther apart. While the first type of perturbation leaves the unit cell of graphene lattice intact, the latter two result in a bigger unit cell (a hexamer of atoms) thus folding the Dirac bands above the light line. A complete theory of all these perturbations combined in one system described by the tight-binding model and the derivation of corresponding effective Hamiltonians is presented in Supplementary Section A. Thus, while valley-Hall photonic system is non-radiative, the modes of the system with Wu-Hu and Kekulé type of perturbations are radiative (leaky), allowing far-field probing of the modes[18-20] and even generation of vortex beams (both for mass profiles without[39,40] and with vortex singularity[28-31]).

The system of interest with all three types of perturbations can be described by an effective Hamiltonian of the form

$$\hat{\mathcal{H}} = \hat{\Gamma}_1 k_x + \hat{\Gamma}_2 k_y - \hat{\Gamma}_3 m_{VH} - \hat{\Gamma}_4 m_{SH} - \hat{\Gamma}_5 m_{KK} \tag{1}$$

where $\hat{\Gamma}_i$ (with $i = 1,2,3,4,5$) is a set of 4x4 matrices that span our synthetic Hilbers space, $\mathbf{k} = \{k_y, k_y\}$ is wavenumber, and $m_{VH}, m_{SH}$, and $m_{KK}$ are valley-Hall, spin-Hall, and Kekulé masses, respectively.

Variations across space of any of the three masses can be employed to obtain nonzero winding in 2D, which is illustrated by three greater circles on an abstract mass sphere $m^2 = m_{VH}^2 + m_{SH}^2 + m_{KK}^2 = const$ in Fig. 1a: (i) spin-Hall/valley-Hall winding ($m_{KK} = 0$) – horizontal orange circle, (ii) Kekulé/valley-Hall winding ($m_{SH} = 0$) – vertical magenta circle, (iii) an arbitrary combination of the three masses - titled greater circle in black dashed line. It is evident that any evolution along any path on a mass sphere enclosing a solid angle of $2\pi$ will give rise to the geometrical phase $\phi_G = \pi$, as confirmed experimentally in[41]. In the context of this work, however, it is important to realize that any such path will result in winding (topological charge of the vortex) of $Q_T = \phi_G/\pi = 1$. Indeed, such vortices, with continuous winding in Kekulé and spin-Hall[28] and in valley-Hall and spin-Hall masses[30], have been demonstrated and used to create topological defect lasers, and, with discrete modulation, to trap zero-dimensional topological polaritons[31].

In what follows, we will only work with vortex structures based on valley-Hall and spin-Hall masses $m_{VH}, m_{SH}$, and we set $m_{KK} = 0$, as in Ref.[43] (see also Supplementary Section B). As demonstrated below, this allows us to achieve control over the radiation of the modes by having regions where the radiation should be suppressed to be dominated by valley-Hall masses, and where the leakage should be maximized, the maximal value spin-Hall mass (and vanishing valley-Hall mass) is employed.

The set of transformations that generate 1D (line defect) and 2D (area/bulk defect) freeform topological resonators is depicted in Fig. 1(c,d,e). Here the vector mass field $\boldsymbol{m} = \{m_{SH}, m_{VH}\}$, with $m = |\boldsymbol{m}| = m_0$, is depicted by arrow plot with arrow color-coded by the mass angle $\tan(\phi) = m_{VH}/m_{SH}$. Thus, horizontal arrows correspond to the purely spin-Hall type perturbations (shrinking and expanding of hexamers for $\phi = 0$ and $\phi = \pi$, respectively), while vertical arrows to pure valley-Hall structured (with two dimerization possibilities – frequency staggering between the two sublattices $a$ and $b$, $\epsilon_a > \epsilon_b$ for $\phi = \pi/2$ and $\epsilon_a < \epsilon_b$ for $\phi = 3\pi/2$ – as detailed in Supplementary Information and in Ref.[41]).

Starting with the conventional vortex carrying a full topological charge, we split it into two semi-vortices which are moved apart (in any direction), producing a topological line defect resembling a Dirac string that is terminated by two half-integer magnetic monopole charges. The inspection of the mass-field distribution across this line in Fig. 1(b) shows that this is a pure spin-Hall domain wall with $m_{SH} = -m_0$ in the top (shrunken) domain and $m_{SH} = m_0$ in the bottom (expanded) domain. Similarly, the topological line defect formed by deforming the vortex into a vertical line yields a pure valley-Hall interface between $m_{VH} = -m_0$ on the left and $m_{VH} = m_0$ on the right. Any other orientation will yield a hybrid domain wall with mixed spin/valley-Hall masses. Regardless, any of such domain walls host boundary modes that are gapless and one-way polarized (pseudo-spin, valley, or a mix) according to their respective mass angle[41].

These domain walls in our case are terminated by semi-vortices and, thus, form finite-length resonators (strings) - for the boundary modes, and are expected to be quantized according to the length and boundary conditions at the ends of the string. Our analytical calculations for the Dirac model (1) (Supplementary Section C) show that the reflection phase from the ends vanishes, yielding the quantization condition (constructive self-interference $kL = \pi N$, $N = \dots, -2, -1, 0, 1, 2, \dots$). Due to the linear $\epsilon = k$ (Dirac velocity is set to $v_D = 1$ for simplicity here) and gapless character of the modes, it then becomes evident that, amongst equally spaced modes with spectral range of $\pi/L$, there will always be a spectrally stable mid-gap "zero-energy" mode corresponding to the *phaseless* mode $k = 0 = \epsilon$, that will remain pinned to the same energy regardless of the length of the string.

Returning to our original picture of vorticity allows us to give a complementary understanding of this zero-energy mode in terms of winding. Thus, the vortex cavity mode upon the transformations from Fig. 1b to Fig. 1c in turn changes into the *phaseless* interface mode trapped to the string with the energy evenly distributed along its length.

Another transformation can then be applied to the topological string in Fig. 1c by deforming it into a finite domain Fig 1d by stretching. This transformation yields a gapless ($m = 0$) region surrounded by the gapped domain with the mass field $\boldsymbol{m}$ winding around it. It is evident that, similar to the string case, this bulk Dirac cavity will support a set of modes formed by the modes propagating in the gapless legion and reflected from its boundaries. However, in contrast to the case when no winding is present[36], the modes are guaranteed to possess the zero-energy mode which is pinned to the same frequency regardless of the cavity size. The zero-energy mode gain in this case is expected to have constant phase within the cavity due to the $k = 0 = \epsilon$ condition.

The above results clearly show that the zero-energy state has topological origin and enables a new mechanism for spectrally stable trapping with the uniform distribution of energy and constant phase, which can be vital for many applications where such the "phase locking" between different regions of the resonator is important.

**Numerical and experimental demonstrations**

To confirm the existence of topological modes described by the abstract Dirac modes above in real photonic systems, we first carried out full-wave electromagnetic modelling to observe the morphing of the vortex mode into a string (line defect). The structure emulating the Dirac physics is a triangular array of hexamers formed by equilateral triangles. The perturbations to be applied here represent (i) valley-Hall perturbation (detuning sized of triangles) giving rise to $m_{VH} \neq 0$ and (ii) pseudo-spin-Hall perturbation (shrinking/expansion) leading to $m_{SH} \neq 0$. All the structure parameters, perturbations to the lattice, are optimization protocol to ensure the fixed mid-gap frequency position for any value of perturbations are identical to those reported by Kawaguchi[41].

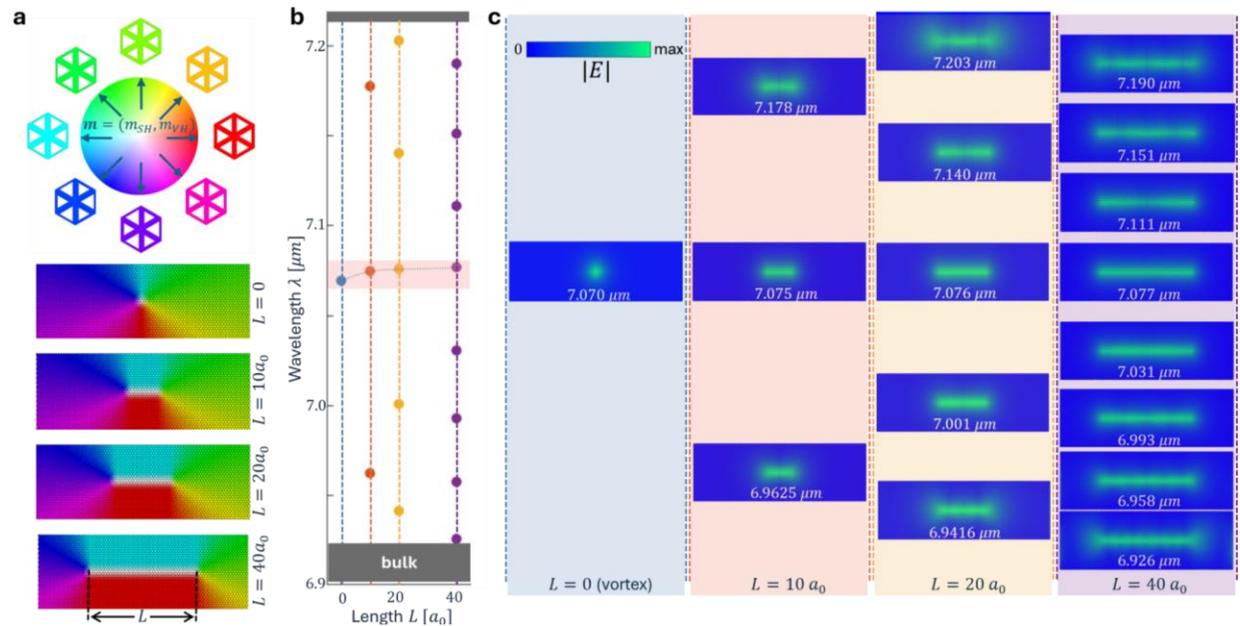

**Fig. 2. Line-defect topological cavity: spectra and near-field mode profiles. a**, Vortex-like modulation of the Dirac mass, encoded in phase color and unit cell design. **b**, Simulated spectra and **c**, electric field magnitude profiles of bound states in line defects of varying lengths $L$.

Examples of the unit cells (with the color-coded continuum) with different perturbations applied (and the resultant orientations of the mass field vector in spin-Hall/valley-Hall plane) are shown in Fig. 1a, top panel, along with four different mass field profiles varying from vortex to line string of increasing length ($L = 10a_0$, $20a_0$, and $40a_0$, where $a_0$ is the lattice constants) shown in Fig. 1a, bottom panel. The results of the eigenfrequency modeling shown in Fig. 1b demonstrate the evolution of the wavelengths $\lambda$ of the modes of the structures in Fig. 2a, with the mid-gap mode present for all values of the cavity lengths. As expected, an increase in cavity size (string length) gives rise to an increased number of resonant modes. The field profiles in Fig. 2b confirm that the mid-gap mode does not possess any nodes, with a uniform phase reminiscent of zero-index photonics[37]. Indeed, the field does not exhibit any variation between unit cells evidencing the phase-less character of the modes, with the only variation present due to the Bloch-Floquet structure of the modes (e.g., $\mathbf{E}(\mathbf{r}) = \mathbf{U}(\mathbf{r}) \exp(i\mathbf{k}_\parallel \cdot \mathbf{r})$ with the wavenumber along the string $\mathbf{k}_\parallel = 0$). Thus, all the unit cells in the resonator oscillate in phase for the mid-gap mode. This is clearly not the case for all other modes, as evidenced by visible variations in the field and the presence of standing-waves and associated nodes. The spectral position exhibits noticeable stability with respect to the variation in the string length, with the most of the shift (5nm, or 0.07%) taking place for the transition from the point-like vortex defect to the short ($L = 10a_0$) string, which then asymptotically approaches $\lambda = 7.077$ μm.

To experimentally observe the transition from point vortex to the linear sting-vortex defect, a set of samples was fabricated in a $1\mu m$-thick silicon on sapphire substrates. The patterns were made by standard e-beam lithography followed by the inductively couple plasma etching (see detail in Methods section). The structures were designed to host the modes of interest near the mid-IR wavelength of 7μm where a custom-built spectroscopy suite (based on a quantum cascade laser and an infrared camera – as detailed in Methods) was used to carry out both polarization resolver and unpolarized imaging of the structures, including in the momentum space (Fourier plane imaging). The collected reflectivity spectra for three selected string lengths, $L = 0$ (vortex), $L = 20a_0$, and $L = 30a_0$, are shown in Fig. 3a. The presence of peaks in the spectra confirm the presence of modes in the region of interest with the number of modes increasing proportionally to the sting length. The modes are also clearly visible in the images in both real and momentum spaces (Fig. 3b,c). As expected, the transition from the point to increasingly longer string leads to longer trapping region, with noticeable drop in intensity due to the even energy spreading in the cavity in real space, while in the momentum space, the lesser localization of the modes in real space leads to increasingly sharper image in momentum space with the momentum $k$ of the modes concentrated predominantly around zero (normal reflection), which is shown in Fig. 3**b** for the mid-gap modes.

To contrast the nature of the zero-energy string-vortex mode and other, higher-order standing wave modes, we carried out polarization resolved imaging of the far field of the structure at resonant frequencies. The relative position of all observed resonances for the structure with $L = 20a_0$ are shown on the left panel in Fig. 3c. The resultant field profiles for different orientations of the polarizer $\theta_p$ reveal that for the mid-gap and the first-order mode shown next to it clearly reveal very different behavior: while for the first-order mode the mode is always visible (with noticeable redistribution of visible intensity over the string length for different values of $\theta_p$), the mid-gap mode disappears for $\theta_p = 90$ (vertical orientation of polarizer). This indicates that the mid-gap mode is linearly polarized along the string, as depicted in Fig.3d, where the reconstructed polarization evolution along the string (averaged along y-axis) is plotted on top of the field profiles for the two modes. The linear polarization of the mid-gap mode can be readily explained by its phase-less nature, thus confirming it. Based on the pseudo-spin-polarized edge-mode interpretation of the modes hosted by the string, all the modes represent a superposition of the forward and backward propagating modes which are known to have opposite handedness, where no total phase is acquired upon reflections from the string ends (Supplementary Section C), thus yielding the four-component wavefunction $u(x) \sim s_\uparrow \exp(ik_\parallel x) + i\, s_\downarrow \exp(-ik_\parallel x)$. The phase-less $k_\parallel = 0$ character of the mid-gap mode therefore leads to $u(x) \sim s_\uparrow + is_\downarrow \sim \{1, -i, i, -1\}$ in the spin-Hall basis of circularly-polarized states, where the second and fourth components are the amplitudes of dipoles responsible for radiation; their superposition thus produces linear polarization. This explains the uniform linearly polarized far-field originating from the vortex-string resonance. Similar considerations explain a more complex and position-dependent phase and rotating polarization profiles of higher order modes $u(x) \sim \{i\sin(k_\parallel x), \cos(k_\parallel x), \cos(k_\parallel x), -i\sin(k_\parallel x)\}$, with $k_\parallel = \pi N/L$.

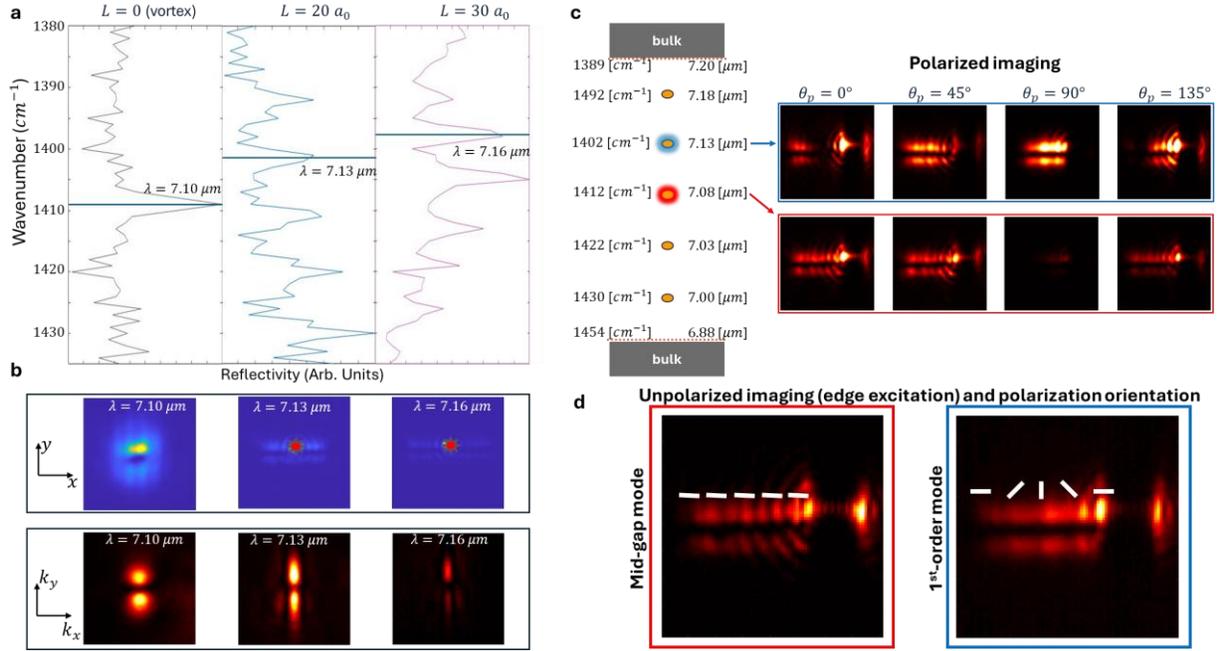

**Fig. 3. Experimental characterization of field confinement, radiation, and polarization in line-defect modes for the spin-Hall orientation of the string. a,** Measured reflectivity spectra together with **b,** representative experimental images at the mode resonances in real and Fourier space with the same intensity scale. **c,** Experimentally extracted relative positions of resonances and the polarization resolved far-field profiles for two selected (zero-energy and first-order) modes **d,** The polarization state of the leaked far-field along the string: uniform linear for the mid-gap mode (red-framed) and rotated for the first-order mode (blue-framed) on top of the unpolarized real-space images.

The topological nature of the mid-gap vortex sting resonance endows with some degree of protection. We first explore its protection against global gauge transformation that gives rise to the rotation of the mass field $\boldsymbol{m}$ over the entire structure, $\phi(\mathbf{r}) \to \phi(\mathbf{r}) + \Phi$, by a constant angle $\Phi$. Such transformation apparently does not affect winding and therefore should keep the spectral position intact, and, for the case in Fig. 1b is equivalent to the rotation of the string by the same angle. We note that in general rotations do not have to be restricted to the valley-Hall-spin-Hall plane we consider here, as for any even out of plane rotations the winding in Fig. 1a is preserved.

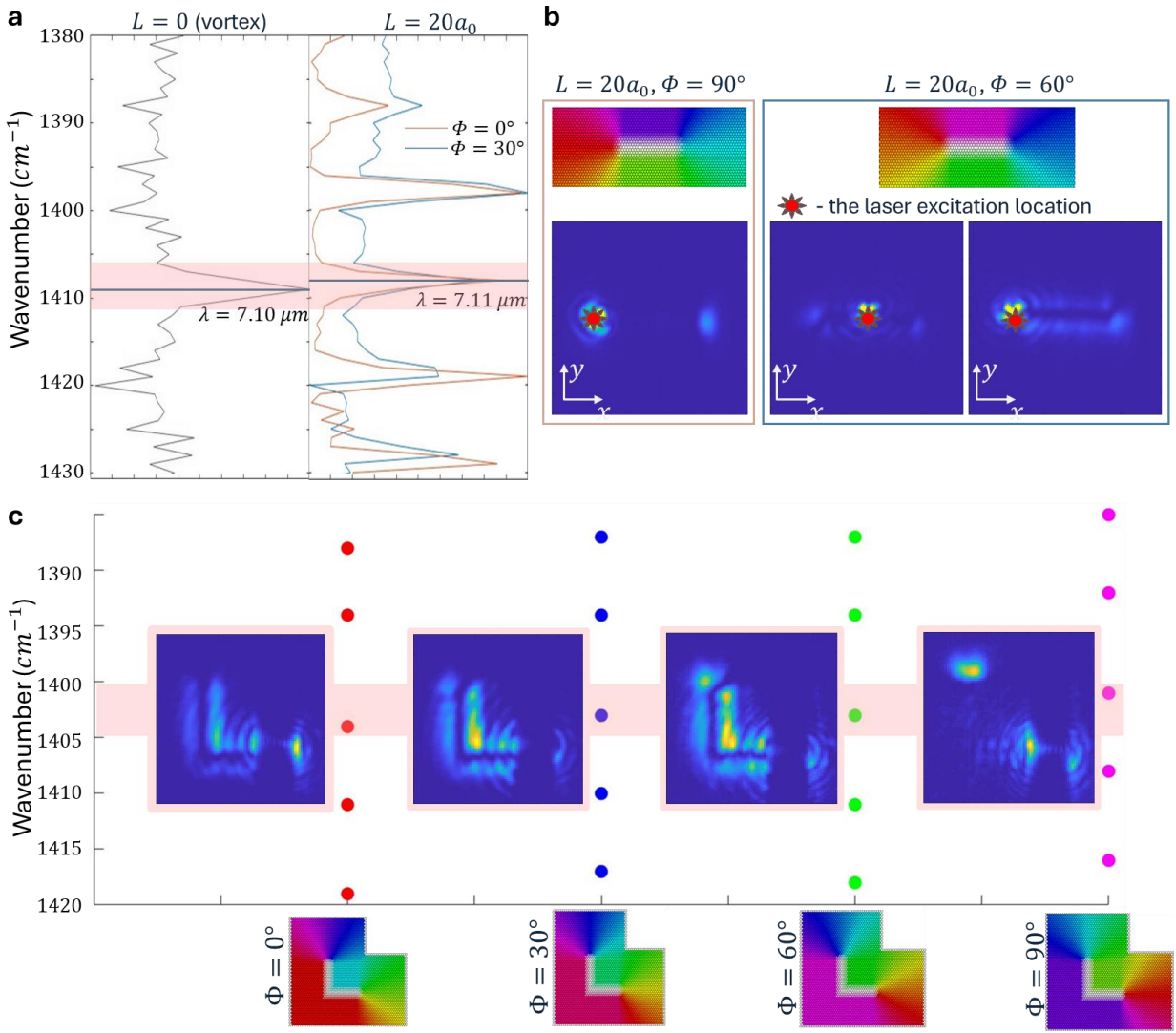

**Fig. 4. Phase-dependent reshaping of radiation and bending-resilience. a**, Reflectivity spectra and **b**, real-space images for line defects with vortices rotated by angle $\Phi$, showing the $\Phi$-dependence of radiative properties (excitation spots marked with stars). **c**, Spectra and mode profiles for corner-shaped defects with different global phase-rotation.

However, for the realistic photonic structures, the rotation in the valley-Hall-spin-Hall plane brings new interesting opportunities for radiative control. Indeed, so far, we considered only the strings corresponding to the spin-Hall domain walls (between shrunken and expanded regions). This choice was justified by the fact that this type of modes is the most radiative which makes their imaging more straightforward. In contrast to this, the valley-Hall interface would be completely non-radiative, while any hybrid class of boundary modes[41] would have their radiative properties defined by the orientation of the string in Fig. 1b, or, for the globally transformed structure, by the rotation angle $\Phi$.

Figures 4a,b illustrates the effect of the global rotation $\Phi = 90°$ on the mid-gap resonance, converting the maximally radiative spin-Hall-type mid-gap mode into the completely nonradiative valley-Hall type mode with radiation from the interface fully suppressed. In fact, the only way to excite this mode was to focus the laser on one of the end points. The mode imaging in real space, shown in Fig. 4b left panel, indeed confirms the radiative leakage only from the tails. A reduction of the global rotation to $\Phi = 60°$, while leaving spectral positions all the resonances (mid-gap and higher-order) nearly intact, does make the mode weakly radiative with no visible far-field profiles for excitation at the center and (still more efficient) at the tails of the string. Thus, the proposed resonators not only offer spectral stability and phase locking in the resonator, but also the possibility to continuously tune radiative quality factor of the mode.

Another class of robustness that we can expect for the system in hands is related to the fact that the edge modes that live on the string exhibit resilience with respect to sharp bending, especially when the interface is adiabatic[42]. In our case, there is always one unperturbed unit cell originating from the initial origin of the vortex point in Fig. 1a, which should make our modes more robust to bending. To confirm such robustness, we fabricated a set of samples where the string is bend by 90 degrees – the most disruptive type of bending not commensurate with the symmetry of the lattice – and tested how such resonators perform for different values of the global rotation $\Phi = 0°, 30°, 60°, 90°$, as shown in Fig. 4c. The spectral position of the modes remains nearly unchanged with the shift of less than 4 cm$^{-1}$ which is attributed to the slight random difference in the patterns due to non-idealities of the nanofabrication (all modes shit nearly equally). At the same time, the radiative pattern is very different, with the least radiative resonator having valley-Hall cladding (dimerized latices) across the bent string with its tales terminated by spin-Hall masses (shrunken/expanded hexamers latices).

As a next step, we then investigate photonic vortex structures morphing into the bulk vortex defects resonators – the vortex 2-brane. The results of numerical modelling, presented in Fig. 5a, confirm that the massless $m = 0$ Dirac 2-brane region supports several modes. Among others, there is mid-gap ("zero-energy") mode with uniform field distribution evidencing the phase-locked nature of this mode with no variation in Bloch phase. Other modes, in contrast, have field profiles with clear standing-wave patterns, i.e., have nodal lines. Similarly to the case of the sting type defects, these higher-order modes of the 2-brane have their spectral positions dependent on the size of the cavity, while the mid-gap mode demonstrates an incredible spectral robustness with respect to both size and shape of the 2-brane. The numerical results in 5a for three different shapes (circular, rectangular, and triangular) of the Dirac cavity clearly reveal this remarkable spectral stability of the mid-gap mode with respect to the cavity shape.

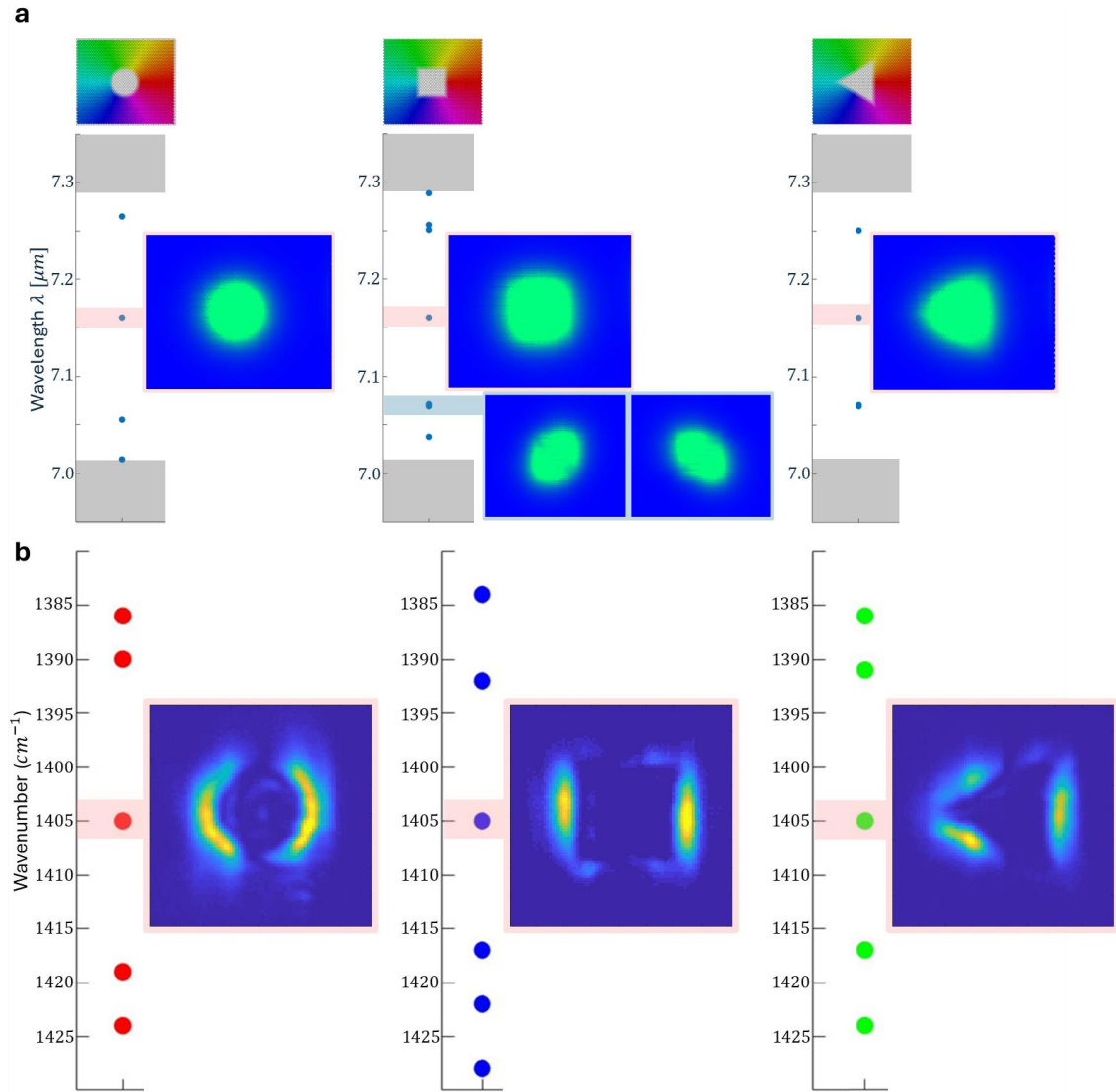

**Fig. 5. Free-form 2D resonators – topological vortex 2-branes**. **a**, Numerically calculated spectra and mode profiles for circular, square, and triangular cavities with massless $m = 0$ (unperturbed geometry) Dirac interiors surrounded by a mass vortex exterior: the field magnitude inside the gapless region is uniform. **b**, Measured spectral positions of the experimentally measured resonances and real-space camera images of the mid-gap modes acquired from the far-field.

To experimentally confirm such robustness, we have prepared three samples with resonators of three different shapes and carried out far-field spectroscopy, exciting the cavity by the mid-IR laser beam focused right outside on the edge of the cavity where the system is most radiative, i.e., where the mass vector is of purely spin-Hall type ($\mathbf{m} = \{\pm 1, 0\}$). The resultant spectral positions of the resonances for the three cases confirm predicted spectral stability of the mid-gap mode, which remain at the same spectral position regardless of the shape of the 2-brane. We note that field trapped in the gapless region is non-radiative (does not have propagating diffractive orders) and therefore cannot be seen in the far-field images, rendering the images with bright spots from the most radiatively leaky $m_{SH} = \pm 1$ regions of the cladding, which ensures long radiative lifetime of these modes.

**Summary**

We have proposed and experimentally demonstrated a class of free-form 1D and 2D spectrally stable resonators. Different from prior demonstrations of free-form resonators, our demonstration ensures inherent robustness, rooted in their topological origin – nonzero winding of the vector mass field – of (zero-energy) modes. These modes appear to be phase-free and therefore always possess monopolar profiles with their phase being locked among all unit cells in the photonic crystals in which such resonators are implemented. When realized in metasurfaces, the relation between the mass vector orientation and radiative leakage allows control over the radiative properties of such modes, including distribution in the far field, and, potentially, radiative lifetimes of the modes. Such unique characteristics of the proposed vortex resonators, combined with their arbitrary shape and size/volume, make them highly attractive for applications. We envision that such resonators implemented across spectral ranges to enable (i) novel approaches to enhanced light-matter interactions with engineered overlap between the field of optical modes and solid-state systems, (ii) novel approaches to lasing of spatially extended yet spatially coherent modes in active metasurfaces and photonics crystals, and (iii) enhanced nonlinear optical phenomena due to predictable phase relations and added control over field intensity distributions, to name just a few. We are confident that similar resonators can be extended to other physical systems, including acoustic and mechanical platforms modes[43-46], where topological materials analogous to those studied here have garnered significant interest in the research community.